\documentclass[letterpaper, 10 pt, conference]{ieeeconf}  %

\IEEEoverridecommandlockouts                              %
\usepackage{amsmath,amsfonts,bm}

\def\eqref#1{equation~\ref{#1}}
\def\Eqref#1{Equation~\ref{#1}}

\def\1{\bm{1}}

\def\vtheta{{\bm{\theta}}}

\def\vf{{\bm{f}}}
\def\vg{{\bm{g}}}

\def\vp{{\bm{p}}}

\def\vu{{\bm{u}}}

\def\vx{{\bm{x}}}
\def\vy{{\bm{y}}}
\def\vz{{\bm{z}}}

\def\mK{{\bm{K}}}

\def\mP{{\bm{P}}}
\def\mQ{{\bm{Q}}}
\def\mR{{\bm{R}}}
\def\mS{{\bm{S}}}

\def\mU{{\bm{U}}}

\def\mX{{\bm{X}}}

\def\mZ{{\bm{Z}}}

\def\mSigma{{\bm{\Sigma}}}

\DeclareMathAlphabet{\mathsfit}{\encodingdefault}{\sfdefault}{m}{sl}
\SetMathAlphabet{\mathsfit}{bold}{\encodingdefault}{\sfdefault}{bx}{n}

\usepackage{amsmath}
\usepackage{amssymb}
\usepackage{fixmath}

\newcommand{\mur}[3]{\mathbold{\mu}_{\overrightarrow{{#1}_{#3}^{#2}}}}
\newcommand{\mum}[3]{\mathbold{\mu}_{{#1}_{#3}^{#2}}}

\newcommand{\sigr}[3]{\mathbold{\Sigma}_{\overrightarrow{#1}_{#3}^{#2}}}
\newcommand{\sigm}[3]{\mathbold{\Sigma}_{{#1}_{#3}^{#2}}}

\newcommand{\laml}[3]{\mathbold{\Lambda}_{\overleftarrow{#1}_{#3}{#2}}}
\newcommand{\lamr}[3]{\mathbold{\Lambda}_{\overrightarrow{#1}_{#3}{#2}}}
\newcommand{\lamm}[3]{\mathbold{\Lambda}_{{#1}_{#3}^{#2}}}

\newcommand{\mat}[1]{\mathbold{#1}}

\newcommand{\veta}{\mat{\eta}}
\newcommand{\vxi}{\mat{\xi}}

\newcommand{\sigEta}[1]{\mat{\Sigma}_{\veta_{#1}}}
\newcommand{\sigXi}{\mat{\Sigma}_{\vxi}}

\newcommand{\tran}{^\intercal}
\newcommand{\inv}{^{\text{-}1}}

\makeatletter
\newcommand{\mathleft}{\@fleqntrue\@mathmargin0pt}
\newcommand{\mathcenter}{\@fleqnfalse}
\makeatother

\newcommand{\itwoc}{\textsc{i2c}}

\usepackage{etoolbox}
\makeatletter
\patchcmd{\@makecaption}
{\scshape}
{}
{}
{}
\makeatother

\usepackage{times} %

\usepackage{tikz}
\usetikzlibrary{arrows}
\usetikzlibrary{fit}
\usepackage{pgfplots}
\pgfplotsset{compat=newest}
\usepgfplotslibrary{groupplots}
\usepgfplotslibrary{dateplot}

\usepackage{booktabs}
\usepackage{threeparttable}

\usepackage{xcolor}

\definecolor{color0}{rgb}{0.75,0,0.75}
\definecolor{color1}{rgb}{0,0.75,0.75}

\title{\LARGE \bf
Advancing Trajectory Optimization with Approximate Inference: Exploration, Covariance Control and Adaptive Risk
}

\author{
Joe Watson, Jan Peters%
\thanks{The authors are with the Department of Computer Science, Technische Universit\"at Darmstadt, Germany.\newline
Contact \texttt{watson@ias.informatik.tu-darmstadt.de}.}
}

\begin{document}

\maketitle
\thispagestyle{empty}

\begin{abstract}
	Discrete-time stochastic optimal control remains a challenging problem for general, nonlinear systems under significant uncertainty, with practical solvers typically relying on the certainty equivalence assumption, replanning and/or extensive regularization.
	Control as inference is an approach that frames stochastic control as an equivalent inference problem, and has demonstrated desirable qualities over existing methods, namely in exploration and regularization.
	We look specifically at the input inference for control (\itwoc{}) algorithm,
	and derive three key characteristics that enable advanced trajectory optimization: An `expert' linear Gaussian controller that combines the benefits of open-loop optima and closed-loop variance reduction when optimizing for nonlinear systems, inherent adaptive risk sensitivity from the inference formulation, and covariance control functionality with only a minor algorithmic adjustment.
\end{abstract}

\section{INTRODUCTION}
\label{sec:intro}
The control of stochastic environments is a ubiquitous problem across many domains, but remains challenging computationally for the general case.
Stochastic optimal control (SOC) solvers trade off computational complexity, exploitation of domain knowledge, use of simplifying assumptions and/or numerical sensitivity.
In this work, we discuss these trade-offs from the perspective of \emph{control as inference}~\cite{DBLP:conf/aips/KappenGO13, 06-toussaint-ICML}, which seeks to frame stochastic control as a probabilistic inference problem.
This translation from optimization to inference can be seen as a subset of probabilistic numerics~\cite{hennig2015probabilistic}, which utilize statistical methods to solve numerical problems, providing uncertainty quantification, regularization and faster convergence.
These viewpoints are made considering input inference for control (\itwoc{})~\cite{i2corl}, a fully probabilistic inference-based solver that frames SOC as input estimation.
Unlike classical control theory, an inference-based approach naturally lends itself to the manipulation of uncertainties, while also benefiting from mature approximate inference methods for complex dynamics and uncertainties where exact inference is intractable.
Considering nonlinear trajectory optimization, without access to closed-form solutions, a key quality is the ability to iteratively explore in a stable manner.
Many SOC algorithms require regularization heuristics such as line search or trust regions to achieve this, as well as initializing with a sufficiently random solution to encourage progress.
We show inference methods naturally achieve this, using belief akin to a trust region and leveraging an adaptive \textit{risk-seeking} strategy for exploration.
Another numerical issue in nonlinear SOC is optimizing open- or closed-loop. 
While open-loop strategies are brittle, yet simpler to compute, closed-loop controllers offer additional stability and therefore reduce the variance of the state distribution.
However optimizing with feedback tends to yield control-heavy solutions due to the interplay between exploration and local feedback during optimization.
We show how using the belief in the controller during optimization allows for this interplay to be managed, producing superior results on two nonlinear tasks.
Covariance control is also implemented using a minor adjustment to \itwoc{}, due to its similarity to inference, enabling nonlinear distributional control that is simpler compared to alternative approaches.

The paper is structured as follows. Section II discuses the SOC literature and related theory, Section III described control as inference and the \itwoc{} algorithm, and Section III details the extensions to \itwoc{} than enable advanced trajectory optimization. 

\section{STOCHASTIC OPTIMAL CONTROL}
\label{sec:soc}
In this section we outline SOC, baseline methods, the risk-based variant and covariance control.
\subsection{The Finite-Horizon, Discrete-Time Setting}
\label{sec:fhdts}
We specifically consider a stochastic discrete-time fully-observed nonlinear time-varying dynamical system, $\vf_t$, with state $\vx \in \mathbb{R}^{d_x}$ and input $\vu \in \mathbb{R}^{d_u}$, desiring the optimal controls over a time horizon $T$ that minimizes the cost functions $C_{0:T}$ in expectation
\begin{align}
    &\min_{\vu_{0:T\text{-}1}}& &\mathbb{E}[C_T(\vx_T) + \sum\textstyle_{t=0}^{T\text{-}1}\;C_t(\vx_t, \vu_t)]\hspace{1cm}\\
    &\text{s.t.}& &\vx_{t+1} = \vf_t(\vx_t, \vu_t).
\end{align}

From Bellman's dynamic programming perspective, the optimal controls 
can be derived from the value function $V_t$ for the expected cost.
For $t = 0,\dots,T\text{-}1$,
\begin{align}
    V_t(\vx) &= \min_{\vu}\mathbb{E}[C_t(\vx, \vu) + V_{t+1}(\vf_t(\vx, \vu))]\label{eq:value},
\end{align}
defining $V_T$ using the terminal cost.
For time-varying linear dynamics, quadratic costs and Gaussian disturbances, the classic LQG solution can be derived in closed-form from Equation \ref{eq:value}.
However, in practice this value function is intractable globally given arbitrary dynamics, objectives and uncertainties.
As a result, methods involve either devising a means of approximating the value function, or tackling the constrained optimization problem directly instead, ignoring the temporal structure.

Differential dynamic programming (DDP) \cite{jacobson1970differential} exploits the temporal structure of the SOC task, iteratively constructing local Taylor series approximations of the dynamics and cost to compute a local approximate value function for closed-loop optimization.
The stochastic DDP extension (SDDP) to consider the impact of Gaussian disturbances on the expected value functions (SDDP) \cite{5530971}.
To mitigate the computational burden of computing the Hessian, a Gauss-Newton approximation of DDP is used (iLQR \cite{DBLP:conf/icinco/LiT04}, iLQG \cite{todorov2005generalized}).
These methods all require regularization and line-search routines for stable convergence.
Guided policy search (GPS) is a method that uses iLQG to solve a variational form of trajectory optimization~\cite{DBLP:conf/icml/LevineK13}.
This results in a maximum entropy linear Gaussian control law, and a KL divergence constraint on the trajectory distribution is used to stabilize the optimization.

For nonlinear, deterministic systems, sequential quadratic programming (SQP) has been used to compute the optimal action- or state-action sequence when framed as a constrained optimization problem, with quadratic convergence under mild assumptions \cite{bryson2018applied}. 
Linearizing along this trajectory, the LQG solution has then been applied to compute an approximate solution to the SOC problem (T-LQG~\cite{7989080}, T-PFC~\cite{8755450}), justified by a small noise performance bound~\cite{doi:10.1137/0309035}.
While mature and highly optimized SQP solvers can be used off the shelf, they can suffer computationally in the SOC domain.
By failing to exploit the temporal structure of the problem (multiple-shooting) or by considering numerically sensitive objectives (single-shooting), SQPs do not scale gracefully under long planning horizons.

\subsection{Covariance Control}
\label{sec:covariance}
While SOC solvers are typically concerned with optimizing the expected cost over the mean trajectory, methods have also been devised to control higher order statistics.
Covariance control \cite {4048352} specifically looks at constraining the mean and covariance of the terminal state to a target distribution $\vx_T^*$.
The linear Gaussian setting as been extensively studied, for both discrete \cite{8264189} and continuous time \cite{7160692}, where it can be shown that a solution exists should the system be controllable and $\sigm{\vx}{*}{T}{-}\sigEta{T}{>}0$, given process noise covariance $\mSigma_{\veta_t}$.
The hard constraint can be tackled by decomposing the problem into feedforward control for the mean, and linear feedback control for the covariance~\cite{8264189}.
The discrete-time case corresponds to minimizing the relative entropy between two MDPs and minimum-energy LQG \cite{532893, 8264189}, where the terminal cost corresponds to the Lagrange multiplier of the constraint.
The nonlinear Gaussian case has been tackled using stochastic DDP \cite{yi2019nonlinear} and through the combination of sequential convex programs and statistical linearization~\cite{9147505}.
The problem can also be viewed as a form of optimal transport and the Schr{\"o}dinger bridge, which seeks to find the mapping (i.e. dynamical system) that transforms one distribution to another \cite{chen2016modeling}.
The solution here is iterative, using forward and backward Riccati equations until both constraints are satisfied.

\subsection{Risk-Sensitive Control}
\label{sec:risk}
Introduced by Jacobson, risk-sensitive linear exponential quadratic Gaussian (LEQG) control \cite{1100265, whittle1981risk} derives a control law that, unlike LQG, is dependent on the severity of uncertainty in the system dynamics. This sensitivity is determined by the scaling parameter $\sigma$ in the now exponentiated objective,
\begin{align}
    \min_{\vu_{0:T\text{-}1}}
    \text{-}
    \frac{2}{\sigma}
    {\mathbb{E}}
    \left[
    \exp
    \left(
    \text{-}
    \frac{\sigma}{2}
    \left[
    C_T(\vx_T){+}\sum_{t=0}^{T\text{-}1}C_t(\vx_t,\vu_t)\right]
    \right)
    \right],
    \label{eq:risk}
\end{align}
for quadratic costs $C_{0:T}$ and $\sigma\in\mathbb{R}$.
The consequence of this objective transformation is that, while they still form a LQG-like discrete algebraic Riccati equation (DARE), the weights of the value function now also depend on the covariance of the Gaussian disturbance $\veta_t$ and risk sensitivity $\sigma$, resulting in the transformation, 
\begin{align}
    \mP_{\text{LEQG},t} &= (\mP_{\text{LQG},t}\inv+\sigma\sigEta{t})\inv
    ,\\
\text{ given } V_t(\vx_t) &= 
\vx_t\tran\mP_t\vx_t + 2\vp_t\tran\vx_t + p_t.
\end{align}
While the relationship between the linear feedback law and the value function is unchanged from the LQG case, the adjustment of the value function results in `risk-preferring' ($\sigma{>}0$) or `risk-averse' ($\sigma{<}0$) strategies under uncertainty.
On a high level, this can be viewed as a mean / variance trade-off in the evaluated cost.
Moreover, for $\sigma{=}0$ (risk-neutral) the control strategy reduces to the LQG result.  
This behavior has lead to interest in risk-sensitive methods from domains such as quantitative finance and behavioral sciences.
The relationship between dynamics uncertainty and risk is what makes this formulation interesting from the inference perspective, which is discussed in Section \ref{sec:i2c-risk}.

\section{CONTROL AS INFERENCE}
\label{sec:cai}
\begin{figure*}[hbt!]
	\begin{align}
	&\max_{\mX,\mU,\,\vtheta}
	p(\mX, \mU, \mZ, \vtheta)=
	p(\vx_0)
	p(\vz_T|\vx_T,\vtheta)
	\textstyle\prod_{t=0}^{T{\text{-}}1}
	p(\vx_{t+1}|\vx_{t},\vu_t)
	\textstyle\prod_{t=0}^{T{\text{-}}1}
	p(\vz_t|\vx_t,\vu_t,\vtheta)
	\textstyle\prod_{t=0}^{T{\text{-}}1}
	p(\vu_t|\vx_t),
	\label{eq:i2cmodel}
	\end{align}
\vspace{-\belowdisplayskip}
\vspace{-\abovedisplayskip}
    \begin{align}
	&\text{Dynamics} &p(\vx_{t+1}|\vx_{t},\vu_t) &: 
	&\vx_{t+1} &= \vf_t(\vx_{t},\vu_t) + \veta_t,
	&\veta_t &\sim \mathcal{N}(\bm{0}, \sigEta{t}),
	\label{eq:dynamics}\\
	&\text{Cost} &p(\vz_t|\vx_t,\vu_t,\vtheta)&:
	&\vz_t &= \vg_t(\vx_{t},\vu_t) + \vxi_t,
	&\vxi_t &\sim \mathcal{N}(\bm{0}, \sigXi(\vtheta)),
	\label{eq:observation}\end{align}
\vspace{-4\belowdisplayskip}
\end{figure*}

The duality between optimal control and statistical inference techniques dates back to the work of Kalman \cite{kalman1960new} from his work on the LQG problem.
Later, relative entropy (KL divergence) was shown as a means of framing the SOC problem \cite{532893}, also known as `probabilistic control design' \cite{vsindelavr2008stochastic}.
While path integral control \cite{DBLP:conf/aips/KappenGO13} strengthened the connection in continuous-time, adopting methods from probabilistic graphical models demonstrated the relation between message passing for inference and the discrete-time Riccati equations in SOC \cite{toussaint2006probabilistic, hoffmann2017linear}.
AICO \cite{06-toussaint-ICML} demonstrated that, for open-loop nonlinear trajectory optimization, the linearization-based approximate message passing computation resembled Gauss-Newton SOC, but converged faster due to its exploratory forward pass and use of priors over control inputs.
However, AICO's exploration prior was fixed due to its dual role as the control regularization, and the translation of the cost to the distributions was hand tuned.
Input inference for control \cite{i2corl} built on AICO, extending the probabilistic perspective by framing optimal control as input estimation.
The closed-loop controller was defined as the conditional distribution of the state-action trajectory, as in posterior policy iteration (PPI) \cite{rawlik2013stochastic,rawlik2013probabilistic}, and crucially the graphical model was defined such that the controls had independent priors, allowing for iterative, variable exploration that corresponds to maximum entropy control.
Moreover, the translation of the cost function for the graphical model, through a scaling term, was jointly optimized during inference.
The combination of principled exploration and probabilistic regularization enable competitive performance against comparable algorithms for nonlinear optimal control.   
To understand how \itwoc{} works, there are four critical components: The latent variable model, expectation maximization, message passing and the transformation of the control cost functions into likelihoods.

\subsection{Latent Variable Models}
\label{sec:lvm}
Like in state estimation, \itwoc{} uses a sequential latent variable model for the state-action trajectory over time, however in the control as inference setting there is no data.
Instead, the `known' quantity is the desired trajectory $\mZ{=}\vz_{0:T}$, which is some transformation $\vg_t(.)$ of the state-action space.
The transformation is tasks specific, e.g. $\vz{=}[\vx, \vu]\tran$ for LQR, $\vz{=}\vu$ for minimum-energy or $\vz$ is in cartesian coordinates for operation space control for a robot manipulator.
Regardless of form, it is assumed to be fixed and part of the objective.
The inference problem is therefore computing the most likely state-action distribution that reconstructs this desired trajectory, which can be done through optimizing the joint likelihood (\Eqref{eq:i2cmodel}).
As $\mZ$ is designed, rather than sampled data, the belief in \itwoc{} is with respect to \emph{optimality}, rather than typical statistical uncertainty.

\subsection{Cost Functions and Constraints as Likelihoods}
\label{sec:likelihood}
In practice, this desired trajectory $\mZ$ will not be perfectly reconstructed.
Therefore, we augment the observation model with additive uncertainty to account for this mismatch (\Eqref{eq:observation}). 
In Section \ref{sec:fhdts}, we discussed LQR for its tractable quadratic the control costs.
These distances can be expressed more concisely as a Mahalanobis distance,
    $\lVert\vy{-}\vx\rVert_{\mS}^2{=}(\vy{-}\vx)\tran\mS\inv(\vy{-}\vx)$.
Assuming Gaussian distributions, we can draw a connection between LQR and \itwoc{} as the `energy' / log-likelihood of this distribution is also a Mahalanobis distance.
For Gaussian state space models, the log-likelihood takes an attractive quadratic form,
\begin{align}
    \vy &= \vf(\vx) + \bm{\eta}, \hspace{0.7cm}\bm{\eta} \sim \mathcal{N}(\bm{0},\mSigma_{\eta}), \\
    -\mathcal{L}_p(\vy, \vx)
    &= -\log p(\vy, \vx) \notag\\
    &= 
    \lVert\vy-\vf(\vx)\rVert_{\mSigma_{\eta}}^2
    + d_x\log|2\pi\mSigma_{\eta}|.\label{eq:loglike}
\end{align}

However, while the control objective in LQG is affine-invariant, the loglikehood is not.
Therefore, for equivalence with the observation loglikehood, there is an unknown scale factor ($\alpha$) to relate the LQG cost weights ($\mQ,\mR$) to the \textit{precision} (inverse covariance) of the disturbance $\vxi$ in \itwoc{},
$\mSigma_{\vxi}\inv{=}\alpha\,\text{diag}(\mQ,\mR)$.
In Section \ref{sec:i2c-risk}, we discuss how it relates to risk sensitivity.
As the only unknown model parameter in the inference problem ($\vtheta{=}\{\alpha\}$),
a benefit of \itwoc{} is that $\alpha$ can be automatically tuned during inference.

\begin{table*}[t]
\centering
\caption{
The evaluation of SOC algorithms on finite-horizon input-constrained control tasks,
comparing variations based on tackling the stochastic (S) or certainty equivalent (CE) setting and using open-loop (FF), closed-loop (FB) or expert (E) controllers during optimization.
Despite each algorithm using different numerical methods, these features identify similarities in performance. Percentiles were computed from 100 rollouts.
}
\resizebox{\textwidth}{!}{%
\begin{tabular}{llllllllll}
	\bfseries
	Environment & \multicolumn{9}{c}{\bfseries 10th, 90th Cost Percentiles ($\times 10^3$)}\\
	\cmidrule(lr){1-1} \cmidrule(lr){2-10}
                & \itwoc{} (S, E)& \itwoc{} (CE, E)& \itwoc{} (S, FF) & \itwoc{} (S, FB) & \itwoc{} (CE, FF) & \itwoc{} (CE, FB) & T-PFC (CE, FF) & iLQR (CE, FB) & GPS (S, FB)\\
\cmidrule(lr){2-10}
Pendulum & 13.46, 21.53 & \textbf{12.81, 17.11} & 17.72, 21.94 & 19.23, 21.43 & 13.97, 26.77 & 19.49, 22.31 & 18.10, 26.31 & 23.33, 26.46 & 19.45, 20.91\\
Cartpole & 85.06, 87.43 & \textbf{81.83, 83.87} & 89.53, 94.67 & 93.53, 95.71 & 86.93, 89.75 & 121.89, 123.88 & 111.31, 118.57 & 142.23, 145.78 & 120.80, 122.45\\
	\bottomrule
\end{tabular}
}
\vspace{-1.5em}
\end{table*}

\subsection{Inference of the Graphical Model}
\label{sec:inference}
From \Eqref{eq:i2cmodel}, we see that the general \itwoc{} inference problem depends on two unknowns: the latent state-action distribution $\mX,\mU$ and model parameters $\vtheta$.
Performing inference with both these unknown quantities jointly is intractable.
Fortunately, for a Gaussian dynamical system computation can be achieved iteratively using expectation maximization (EM) \cite{dempster1977maximum, Ghahramani96parameterestimation}, which guarantees monotonic improvement.
In the EM convention, the E-Step corresponds to estimating the latent state-action distribution given the model, while the M-Step the optimizes the model parameters $\vtheta$ to maximize the \textit{expected} log-likelihood given the estimated latent distribution.
Using \Eqref{eq:loglike}, this is of the form
\begin{align}
   -\mathbb{E}[\mathcal{L}_p(\vy, \vx)] = 
   &\text{tr}\{\sigEta{~}\inv\mathbb{E}[(\vy{-}\vf(\vx))(\vy{-}\vf(\vx))\tran]\} \notag\\
   &+ d_x\log|2\pi\bm{\Sigma}_{\veta}|.\label{eq:expectedloglike}
\end{align}
This EM procedure is iterated till convergence.
While the M-Step improvement can be expressed in closed-form \cite{i2corl}, the E-Step requires closer consideration.

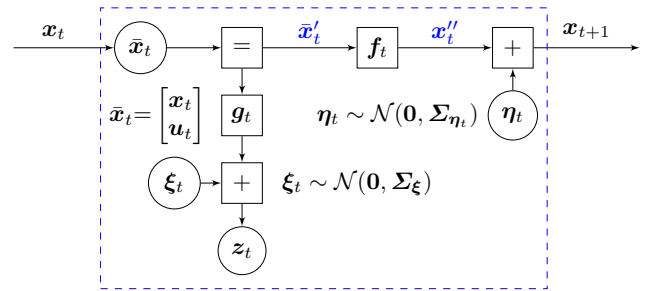
\begin{figure}[b]
	\vspace{.5em}
	\begin{tikzpicture}
[%
 node distance=20mm,
 auto,>=latex',
 box/.style={draw, minimum size=0.6cm},
 short/.style={node distance=10mm},
 transform canvas={scale=0.9} %
]
\node [yshift=0cm, xshift=0cm] (start) {}; %
\node[circle, draw, scale=1.0, right of=start] (joint) {$\bar{\vx}_t$} edge[<-] 
node[above,pos=0.6] {{$\vx_{t}$}}(start);
\node[short, below of=joint] (joint_vec) {\hspace{5mm}$\bar{\vx}_t{=}
	\begin{bmatrix}
	\vx_t \\
	\vu_t
	\end{bmatrix}$};
\node[box, draw, right of=joint, xshift=-0.5cm] (x_fuse) {$=$} edge[<-] (joint);
\node[box, draw, below of=x_fuse, yshift=1cm] (g) {$\vg_t$} edge[<-](x_fuse);
\node[box, draw, below of=g, yshift=1cm] (xi_add)  {$+$} edge[<-] (g);
\node[circle, draw, left of=xi_add, xshift=1cm] (xi) {$\bm{\xi}_t$} edge[->](xi_add);
\node[short, right of=xi_add] (sig_xi) {\hspace{14mm}$\bm{\xi}_t \sim \mathcal{N}(\mathbf{0},\sigXi)$};

\node[circle, draw, below of=xi_add, yshift=1cm] (z) {\hspace{0mm}$\vz_t$} edge[<-] node {} (xi_add);
\node[box, draw, right of=x_fuse, xshift=0cm] (f) {$\vf_t$} edge[<-] node[above=-2pt,pos=0.5] {\textcolor{blue}{$\bar{\vx}_{t}'$}}  (x_fuse); %
\node[box, draw, right of=f,xshift=0.0cm] (eta_add) {$+$} edge[<-]
node[above=-2pt,pos=0.5] {\textcolor{blue}{$\vx_{t}''$}} (f);
\node[circle, draw, below of=eta_add, yshift=1cm] (eta) {$\veta_t$} edge[->] (eta_add);
\node[short, left of=eta] (sig_eta) {\hspace{-14mm}$\veta_t \sim \mathcal{N}(\mathbf{0},\sigEta{t})$};
\node[right of=eta_add] (end) {} edge[<-] node[above] {$\vx_{t+1}$}
  (eta_add); %

\node[box,blue,dashed,inner sep=2mm,fit=(joint)(f)(g)(x_fuse)(xi_add)(z)(eta_add)] {};
\end{tikzpicture}
	\vspace{9em}
	\caption{The factor graph of \itwoc{} for a single timestep.}
	\label{factor_graph}
\end{figure}

To perform the E-Step efficiently, message passing methods are a flexible approach to designing such algorithms~\cite{loeliger2007factor}.
Passing linear Gaussian messages is especially straightforward, following simple rules that can be expressed in closed-form.
For nonlinear Gaussian messages, approximate inference can be performed in several ways.
Akin to the various nonlinear Bayesian filters, Taylor series approximations, quadrature and Monte Carlo methods can be used to approximate the marginalization integral \cite{anderson2012optimal}.
As the graph is a Gaussian dynamical system, the E-Step relates closely to general recursive Bayesian smoothing, using Equations \ref{eq:dynamics} and \ref{eq:observation} as the Gaussian dynamics and `observation' model respectively.
However, the key separation from typical smoothing is the inclusion of the input as part of the probabilistic state. 
With a prior on the state-action distribution at time $t$,
during the forward message passing we compute
$p(\overrightarrow{\vx}_t, \overrightarrow{\vu}_t)$
from the incoming $\overrightarrow{\vx}_t$ computed in the previous timestep by marginalizing over the conditional distribution,
\begin{align}
    p(\overrightarrow{\vu}_t) &=
    \int p(\vu_t|\vx_t{=}\overrightarrow{\vx}_t)p(\overrightarrow{\vx}_t)d\overrightarrow{\vx}_t,\\
    \mur{\vu}{}{t} &= \mum{\vu}{}{t} + \mK_t(\mur{\vx}{}{t}-\mum{\vx}{}{t}),\\
    \sigr{\vu\vu}{}{t} &= \sigm{\vu}{}{t} - \sigm{\vu\vx}{}{t}\sigm{\vx\vx}{}{t}\inv\sigm{\vu\vx}{}{t}\tran
    + \mK_t \sigr{\vx}{}{t} \mK_t\tran,\\
    \sigr{\vu\vx}{}{t} &= \sigm{\vu\vx}{}{t},\;\;
    \mK_t = \sigm{\vu\vx}{}{t}\sigm{\vx\vx}{}{t}\inv.
\end{align}
Where $\mK_t$ is the \itwoc{} feedback control gain, which has been shown to be equivalent to maximum entropy LQR in the deterministic setting \cite{i2corl}.
Note that if the state and action are independent (i.e. $\sigm{ux}{}{t}{=}\bm{0}$) or the state distribution is unchanged ($\overrightarrow{\vx}_t{=}\vx_t$), then no `feedback' is applied and the input distribution remains as the prior $\overrightarrow{\vu}_t{=}\vu_t$.
This feedback term also exhibits the `turn-off phenomena' from dual control \cite{1100635}, as the gain is attenutated as the state uncertainty increases.

\section{INFERENCE-BASED ADVANTAGES}
\label{sec:algorithm}
In this section we introduce algorithmic improvements for \itwoc{} and explain existing qualities from the SOC perspective.
\subsection{Optimizing Expert Linear Gaussian Controllers}
\label{sec:numerics}
When performing trajectory optimization, considering the open- or closed-loop setting is a crucial distinction.
The open-loop approach (i.e. using SQPs) is a simpler optimization problem, but may be brittle and is limited to shorter planning horizons and levels of stochasticity.
On the other hand, closed-loop optimization (i.e. DDP) provides extra stability with the local control law, however the numerical procedure has unfortunate side-effect in the nonlinear setting: the feedback fights exploration during optimization, resulting in highly-actuated, and therefore likely sub-optimal, solutions.
As \itwoc{} derives its controller from the joint state-action distribution, both open- and closed-loop optimization can be incorporated into the E-step by considering the independent ($\sigr{\vu\vx}{}{t}{=}\,\bm{0}$) or full joint distribution respectively.

In Table 1 we compare these variations on two stochasic, nonlinear swing-up tasks, looking at open- and closed-loop optimization for both \itwoc{} and baseline SOC solvers, which also vary between considering the actual stochastic problem or a certainty equivalent approximation.
For both tasks, open-loop methods resulted in better optima but were also high variance in the cost, while the closed-loop alternatives had much lower variance but sub-optimal performance on the simulated systems due to their over-actuation.
Reassuringly, the results of the \itwoc{} variant and equivalent baseline solver were generally similar due to the comparable computation.

These results then suggest that ideally we should be able to combine the benefits of open- and closed-loop optimization to improve performance during inference.
With \itwoc{} this is indeed possible, by adjusting the conditional distribution ($\pi_t$) to act as an `expert' controller, which utilizes the probability that the local controller applies to the current state $p(\vx_t{=}\vx)$,
\begin{align}
	\pi_t(\vu|\vx) &= p(\vx_t{=}\vx)\,p(\vu_t\mid\vx_t{=}\vx) + p(\vx_t{\neq}\vx)\,p(\vu_t).   
\end{align}
This weighting softens the controller to fall back to the open-loop controls $p(\vu_t)$ as $\vx$ moves away from estimated optimal state trajectory.
Therefore, the feedback effect is reduced during significant exploration in the E-step, and as the open-loop controls are well regularized this avoids highly actuated trajectories forming.
Table 1 demonstrates the effectiveness of this addition, where this expert controller matches the open-loop optima but with the closed-loop variance reduction.
For a Gaussian distribution, $p(\vx_t{=}\vx)$ is defined as the confidence interval $\lVert\vx-\mum{\vx}{}{t}\rVert_{\mSigma_{\vx_t}}^2\leq\mathcal{X}^2_k(p)$ for probability $p$, where $\mathcal{X}^2_k$ is the chi squared distribution.
Another interesting aspect of the expert controller results is how certainty equivalence outperforms optimizing the true stochastic setting.
In practice, we also found the stochastic setting required greater care regarding hyperparameter tuning. 
This is because under greater uncertainty the inaccuracy in the approximate inference is more severe, thus requiring greater regularization during inference.
Moreover, from the probabilistic numerics view, greater uncertainty corresponds to greater numerical regularization e.g. the turn-off phenomena in the feedback gains, which may also be at play.
Studying this phenomena is greater detail, especially with more accurate approximate inference techniques, will be a topic of future work.

\subsection{Adaptive Risk-Sensitivity}
\label{sec:i2c-risk}

For Gaussian \itwoc{}, the expected likelihood in \Eqref{eq:i2cmodel} may be reformulated to expose the quadratic cost function from the observation likelihood,
\mathleft
\begin{equation}
    {\mathbb{E}}\hspace{-1mm}\left[
    {A({\mX}{,}{\mU}{,}\alpha)}
    {\exp}
    {\left(
    \text{-}
    {\frac{\alpha}{2}} %
    \left[
    C_T(\vx_T){+}\sum_{t=0}^{T\text{-}1}C_t(\vx_t{,}\vu_t)
    \right]
    \right)}
    \right]
\end{equation}
\mathcenter
where $A(.)$ contains the remaining likelihood and normalizing terms.
Comparing this form to \Eqref{eq:risk}, $\alpha$ may be interpreted as equivalent to risk sensitivity.
For both LEQG and \itwoc{}, $\sigma{=}\alpha{=}0$ corresponds to the deterministic LQR result.
However, for \itwoc{}, $\alpha>0$ due to the probabilistic treatment, resulting in only risk-seeking behavior possible, with risk neural control as the limit of $\alpha{\rightarrow}0$.
Risk-seeking behavior is an inherent issue for control as inference methods, due to the probabilistic formulation naturally attenuating the effect of unlikely trajectories in the objective.
Related methods such as GPS and path integral control also share this property.
However, due to the other terms at play in \itwoc{}, $\alpha$ and $\sigma$ do not have a direct correspondence, nor do their resultant trajectories exactly match.
This discrepancy can be identified by examining the linear Gaussian message passing equations.
First note in the deterministic setting (without $\bm{\eta}_t$),
\begin{align}
\lamm{\vx}{''}{t} &= \lamm{\vx}{}{t+1}, \text{where 
	$\vx_t{''}$ denotes the state after $\vf$.}
\intertext{Moreover in the LQR equivalent case, where there are sufficiently uninformative priors placed on $\mU_t$,}
\lamm{\vx}{}{t+1} &= \lamr{\vx}{}{t+1}{+}\laml{\vx}{}{t+1} \approx \laml{\vx}{}{t+1}\;\text{for large $\sigr{\vx}{}{t+1}$}.
\end{align}
This provides an inference-based perspective why LQR can be solved without considering the forward propagation. 
However, considering disturbance $\bm{\eta}_t$,
\begin{align}
\sigm{\vx}{}{t+1} &= \sigm{\vx}{''}{t} + \sigEta{t},\;\;
\lamm{\vx}{}{t+1} = (\lamm{\vx}{''}{t}\inv + \sigEta{t})\inv,\\
\lamm{\vx}{''}{t}\inv &= (\lamr{\vx}{''}{t}{+}\laml{\vx}{''}{t})\inv,\\
 &=
\laml{\vx}{''}{t}\inv{-}\laml{\vx}{''}{t}\inv (\lamr{\vx}{''}{t}\inv{+}\laml{\vx}{''}{t}\inv)\inv\laml{\vx}{''}{t}\inv,\\
\lamm{\vx}{}{t+1} &\approx (\laml{\vx}{''}{t}\inv + \hat{\bm{\Sigma}}_{\veta_t})\inv.
\end{align}

Therefore, LEQG's risk sensitivity may be viewed as a means to approximate the prior state distribution while using LQR's Riccati equation, through the transformed disturbance term $\hat{\bm{\Sigma}}_{\veta_t}{=}\sigma\sigEta{t}$.  
Moreover, we can see that setting $\sigma>0$ (i.e. risk avoidance) acts to reduce the uncertainty in the trajectory in a way that probabilistically invalid, as $\hat{\bm{\Sigma}}_{\veta_t}$ must be positive definite. 

While risk-seeking behavior is typically seldom desired, $\alpha$'s influence on both exploration and risk suggests that risk-seeking control is useful (for nonlinear systems) by greedily seeking the desired trajectory.
Moreover, further consideration of this term may improve the robustness of the controller by limiting risky strategies.
\subsection{Covariance Control as Inference}
\begin{figure}[!tb]
	\begin{minipage}[t][0.8cm][t]{\columnwidth}
		\hspace{1.4cm}
		\begin{tikzpicture}

\definecolor{color0}{rgb}{0,0.75,0.75}

\begin{axis}[
hide axis,
width=8.5cm,
xmin=10, xmax=50,
ymin=0, ymax=1.0,
legend cell align={center},
legend columns=4,
legend style={/tikz/every even column/.append style={column sep=0.3cm}, draw=none},
]
]

\addlegendimage{semithick, black, mark=x, mark size=3, mark options={solid}, only marks}
\addlegendentry{$\mathbf{x}_0$};
\addlegendimage{semithick, red, mark=x, mark size=3, mark options={solid}, only marks}
\addlegendentry{$\mathbf{x}_g$};
\addlegendimage{semithick, green!50.0!black, mark=*, mark size=1, mark options={solid}}
\addlegendentry{Closed-loop};
\addlegendimage{semithick, color0, opacity=0.5, dashed, mark=*, mark size=1}
\addlegendentry{Rollouts};
\end{axis}

\end{tikzpicture}
	\end{minipage}
	\begin{minipage}[t][5.8cm][t]{\columnwidth}
		\centering
		\input{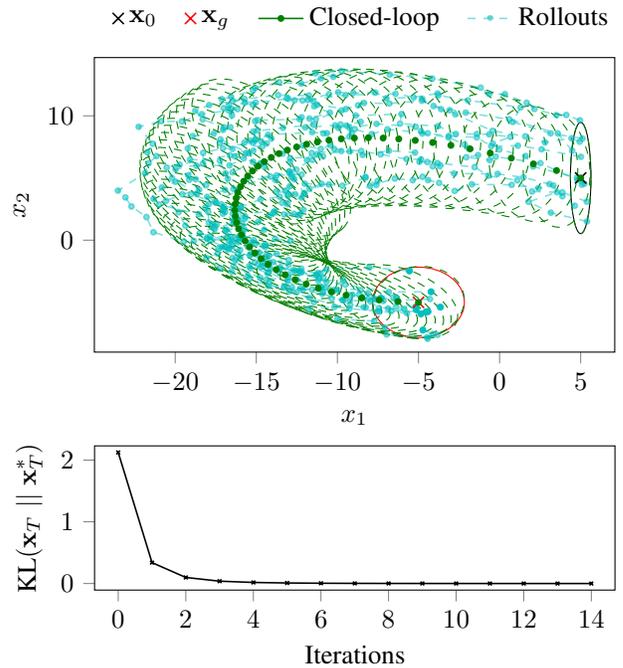}
	\end{minipage}
	\caption{\itwoc{} for minimum-energy exact linear Gaussian covariance control on an unstable system, with a fixed small $\alpha$ and $\sigEta{t}{=}\text{diag}(0.1, 0.1)$.
		The KL divergence is between the terminal goal and closed-loop distributions.}
	\label{fig:covariance}
	\vspace{-1em}
\end{figure}

\begin{figure}[!tb]
	\caption{Nonlinear minimum-energy covariance control on the pendulum swing-up task, using \itwoc{} with approximate inference. Plot depicts the inferred trajectory~\ref{plan} for target distribution \ref{target}, with simulated rollouts \ref{rollout}.}
	\vspace{.5em}
	\begin{minipage}[t][3cm][t]{\columnwidth}
		\centering
		\input{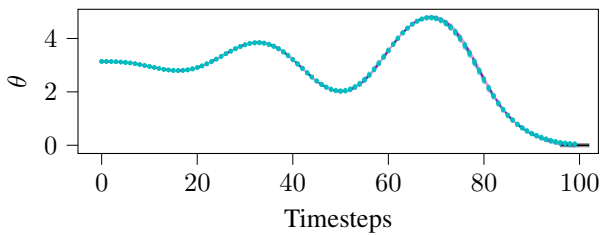}
	\end{minipage}
	\label{fig:cc}
	\vspace{-3em}
\end{figure}

For \itwoc{}, the idea of a terminal cost can be tackled in two ways.
Previously, we have used the idea of a terminal observation function $\vg_T(\vx)$, which allows for arbitrary terminal costs as described in Section \ref{sec:likelihood}.
However, another approach is to work with the terminal latent state distribution.
In other time-series inference applications, e.g. state estimation, the terminal state posterior is set to the prior as there is no additional information to use.
However, for control we can avoid the cost function design and the cost-to-likelihood translation by setting the terminal distribution directly, but are now limited to stipulating the desired state directly rather than a transformation (e.g. $\cos\theta{=}1$).
This approach is equivalent to covariance control (Section \ref{sec:covariance}) as, like solving a Schr{\"o}dinger bridge, iterations of forward and backward Riccati equations are performed until the boundary condition are satisfied.
In this setting, $\mSigma_{\vxi_T}$ now acts as the Lagrange multiplier.
Examining the expected log-likelihood (\Eqref{eq:expectedloglike}) term for the terminal state ($\mathcal{L}_{\vz_T}$) for the direct state optimization case $\vz_T{=}\vx_T{+}\vxi_T$, we get
\begin{align}
\mathcal{L}_{\vz_T}
&\propto\lVert\mum{\vz}{}{T}{-}\mum{\vx}{}{T}\rVert_{\sigXi}^2{+} \text{tr}\{\mSigma_{\vxi_T}\inv\sigm{\vx}{~}{T}\}+\text{const.}\end{align}
which is equivalent the LQG correspondence proved in Goldshtein et al. (Equation 41) \cite{8264189}, where $\mSigma_{\vxi_T}\inv$ is the terminal cost / Lagrange multiplier matrix.
However, rather than compute this term, using our probabilistic framework we can set the terminal distribution directly via the posterior of $\vx_T$.
In this case, the inference iteratively seeks to satisfy the boundary conditions on $\vx_0$ and $\vx_T$.
Figure \ref{fig:covariance} demonstrates covariance control on a linear Gaussian system, where inference is exact.
It should be noted that \itwoc{} uses linear Gaussian controllers, whose uncertainty is required to achieve the desired state distribution, whereas the previous literature solves the task using deterministic control laws.
This variation of \itwoc{} naturally translates to nonlinear systems (Figure \ref{fig:cc}), avoiding the complexity of the additional forward sampling required for DDP-based covariance control \cite{yi2019nonlinear}.
However, the terminal boundary constraint still requires a means of being applied in an gradual manner, due to the iterative aspect of the nonlinear optimization.
The terminal state distribution can be shifted from the prior to the constraint by `annealing' \cite{annealing} the prior during inference 
\begin{align}
	p(\vx_T) = p(\vx_T^*)\,p(\overrightarrow{\vx}_T)^{\beta},\; 1 \geq \beta \geq 0,
\end{align}
where $\beta$ is the annealing temperature on the prior from the forward pass.
By decreasing $\beta$ over iterations, $p(\vx_T){\rightarrow}p(\vx_T^*)$ as $\beta{\rightarrow}0$.
This annealing strategy has previously been applied to regularize inference algorithms.

\section{CONCLUSION}
In this work, we have discussed control as inference and the \itwoc{} algorithm from the perspective of stochastic optimal control.
We have analysed \itwoc{} with respect to the risk-sensitive control, highlighting the similarities between LEQG and inference and how \itwoc{} utilizes adaptive risk during EM.
Moreover, by analyzing the numerical consequences of open- and closed-loop optimization with \itwoc{} and baseline solvers, we have motivated an expert linear Gaussian controller that leverages the state belief to balance exploration with stabilizing feedback, which achieved superior results on simulated nonlinear control tasks.
Finally, we demonstrated how covariance control can be implemented with a minor adjustment to \itwoc{}, enabling exact and approximate solutions in the linear and nonlinear setting respectively.

\section{ACKNOWLEDGMENTS}
This project has received funding from the European Union’s Horizon 2020  research and innovation programme under grant agreement No. 713010 (GOAL Robots).
The paper was inspired in part by many insightful discussions with Hany Abdulsamad and Boris Belousov.

\bibliographystyle{IEEEtran}
\bibliography{lib.bib}

\appendix
\noindent
\textbf{Experimental Details}\\
For the experiment reported in Table 1, task parameters are similar to prior work \cite{i2corl}, but with the Cartpole environment now operating at a 4ms timestep for a time horizon of 500. 
For \itwoc{}, cubature quadrature was used for approximate inference \cite{cubature}.
T-PFC was implemented using the transcription method (i.e. $(T{+}1)d_x{+}Td_u$ state space and $T{+}1$ constraints)
.

\addtolength{\textheight}{0cm}   %

\end{document}